\documentclass[aps,prl,preprint]{revtex4}
\usepackage{graphicx} \usepackage{amssymb,amsmath}

\begin{document}

\title{Hydrogen bonding  and coordination in  normal and supercritical
water from X-ray inelastic scattering}

\author  {Patrick   H.-L.  Sit,$^{1}$  Christophe   Bellin,$^{2}$  Bernardo
Barbiellini,$^{3}$     D. Testemale,$^{4}$     J.-L.    Hazemann,$^{4}$
T. Buslaps,$^{5}$ Nicola Marzari,$^{6}$ Abhay Shukla$^{2}$}

\affiliation{$^{1}$Department  of Physics, Massachusetts  Institute of
Technology,   Cambridge,   MA   02139,   USA  \\   $^{2}$Institut   de
Min\'{e}ralogie   et   de    Physique   des   Milieux   Condens\'{e}s,
Universit\'{e} Pierre et Marie Curie,  case 115, 4 place Jussieu 75252
Paris Cedex  05, France  \\ $^{3}$Department of  Physics, Northeastern
University,   Boston,   MA   02115,   USA  \\   $^{4}$Laboratoire   de
Cristallographie, UPR 5031, 26  Avenue des Martyrs, Boite Postale 166,
38042 Grenoble Cedex 9, France \\ $^{5}$European Synchrotron Radiation
Facility, BP  220, F-38043 Grenoble Cedex,  France \\ $^{6}$Department
of  Materials  Science  and  Engineering, Massachusetts  Institute  of
Technology, Cambridge, MA 02139, USA }

\date{\today}
\begin{abstract}
A direct measure of hydrogen bonding in water under conditions ranging
from the normal state to  the supercritical regime is derived from the
Compton scattering of  inelastically-scattered X-rays. First, we show that
a measure of the number of electrons $n_e$ involved in
hydrogen bonding at varying thermodynamic conditions
can be directly obtained from Compton profile differences.  
Then, we use first-principles simulations to provide a connection between 
$n_e$ and the number of hydrogen 
bonds $n_{HB}$.  
Our study shows that over the broad range studied the 
relationship 
between $n_e$ and $n_{HB}$ is linear, allowing for a direct experimental 
measure of 
bonding and coordination in water. In particular, the transition to
supercritical state is characterized by a sharp increase in the number of 
water 
monomers, but also displays a significant number 
of residual dimers and trimers.
\end{abstract}

\maketitle

%
%
The complexity of the phase diagram of water and its relevance to many
physical,  chemical and  biological  processes is  to  a large  extent
related to the presence  of hydrogen bonds \cite{Eisenberg}.  Over the
years, different methods have  been used to investigate the structure and dynamics 
of  water  \cite{Review1};  nevertheless, many  fascinating  questions still remain.   
Hydrogen bonding and coordination in disordered phases, most notably in 
the liquid, are the subject of intense study and the average 
number of hydrogen bonds ($n_{HB}$) in water has recently become a subject of controversy
\cite{Science,Science2}. X-ray absorption
spectroscopy (XAS) \cite{Science} in water and ice indicated that
the liquid contains significantly more broken hydrogen bonds
than previously thought, but these conclusions have been questioned
\cite{Science2,Galli}. 
The degree of persistence of hydrogen bonding in the region beyond the critical 
point is also an interesting question \cite{soper97}. Several
works report on residual hydrogen bonding in supercritical water 
\cite{Review1,JCP_xraman}. 
The number of hydrogen bonds per molecule $n_{HB}$ 
is a concept of some ambiguity since it depends on geometric definitions. 
The connectivity of the hydrogen bonded network has also been recently 
studied in terms of electronic signatures \cite{Artacho}
highlighting the presence of a fast electronic bond dynamics and reorganization.        

In this work, we highlight how Compton scattering  
- i.e. inelastic scattering of photons by electrons \cite{cooper} 
- can actually measure the number of electrons involved in hydrogen bonding, 
and we use this result to characterize structure and bonding in water 
under a variety of thermodynamical conditions. The Compton profile (CP) is given by 

\begin{equation} 
J(p,{\bf e})=\int n({\bf p'})\delta ({\bf p' \cdot e}-p){\bf dp'}, 
\end{equation} 
where {\bf e} is the unit vector along the scattering direction and $n({\bf p'})$ 
is the electron momentum density; here we only consider the spherical average $J(p)$ 
over ${\bf e}$. The unique sensitivity of CPs to valence electrons and to 
chemical bonding (via $n({\bf p'})$) is at variance with diffraction techniques, 
which are principally sensitive to ionic positions.  Earlier work using Compton 
scattering has in fact revealed the quantum nature of the hydrogen bond in 
ice \cite{Isaacs}, and it has also been suggested that CP can be a measure 
of the amount of H-bonding in water 
\cite{Ghanty,ragot,bba02,Hakala1,Hakala2,Hakala3,Nygard}. 
The Compton profile reflects the distribution of electronic 
momentum in the ground state of the measured system. Changes 
in bonding induce changes in this distribution and by tracking 
these changes both in experiment and calculations for water 
we establish two central points.
First, we show that we can measure directly the number of electrons 
associated with change in bonding induced by varying thermodynamic conditions. 
Then, we demonstrate a very robust linear relationship between $n_{HB}$ 
and $n_{e}$, the number of electrons per molecule involved in hydrogen bonding, 
from ambient to supercritical conditions.
Therefore, the measured $n_{e}$ becomes a direct and unambiguous measure of $n_{HB}$.

\begin{figure} \begin{center}
\includegraphics[width=\hsize,width=8.0cm]{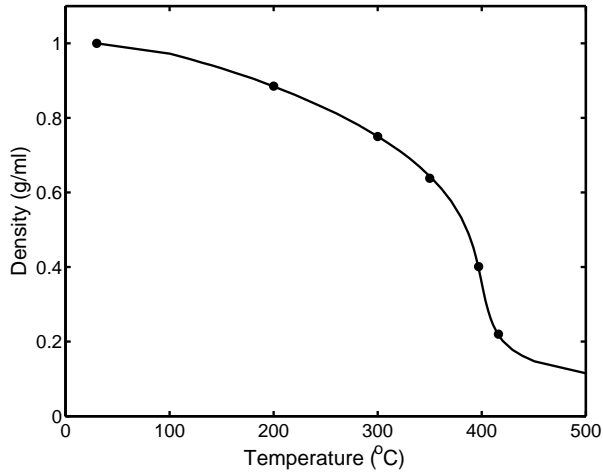}
\end{center}
\caption{Experimental measurements were taken along the
isobaric curve for P=300 bar.}
\label{diag_water} \end{figure}

%
%
As shown in Fig~\ref{diag_water},
our measurements cover the
temperature  range  along  the  isobar  P=300  bar,  and  include  the
following  state  points:   (a)  T  =  30  $^{o}$C   (density  =  1.01
g/cm$^{3}$), liquid  water reference; (b)  T = 200 $^{o}$C  (density =
0.89 g/cm$^{3}$), T = 300 $^{o}$C  (density = 0.76 g/cm$^{3}$) and T =
350 $^{o}$C (density = 0.67 g/cm$^{3}$), all showing precursor effects 
of hydrogen-bond breaking and change of  local tetrahedral structure; (c) 
T = 397
$^{o}$C (density = 0.40  g/cm$^{3}$), critical point neighborhood; (d)
T = 416 $^{o}$C (density = 0.21 g/cm$^{3}$), supercritical conditions.

Our  measurements were  taken  at the  European Synchrotron  Radiation
Facility in Grenoble  at beamline ID15B using the  56 keV experimental
set-up    and     a    specially    designed     high-pressure    cell
\cite{Testemaletech}. The scattering angle  was set at 144.4$^{o}$ and
the  back-scattered  photons  for  the  CP  were  collected  by  a  Ge
multi-element  detector. After  subtracting the  very  weak background
(5$\times$10$^{-3}$\% at the Compton peak),  the raw data were coverted 
from
wavelength scale to  momentum scale, and multiple scattering contributions
were subtracted from the measured profile in order to obtain the total
profile. The multiple scattering contributions were calculated using
Monte-Carlo simulations with each dataset treated separately 
in  order to  account the  change in  density. The  ratio  of multiple
scattering to the total ranges between  0.9\% at 30 $^{o}$C to 5\% at 416 
$^{o}$C.
Our  statistics   ranges  between  2.1$\times$10$^{6}$  counts   at  30 
$^{o}$C  to
1.3$\times$10$^{6}$   counts   at   416 $^{o}$C   at   the   Compton   
peak.    The
energy-dependent resolution function  was obtained from the full-width
at half  maximum of the elastic  peak (0.51 atomic  units of momentum).   
Now, since the  CP  is  a projection  of  the electronic  momentum
density, its integral  yields the total number of electrons and provides 
a convenient normalization. 
In  this  work,   we  focus  our  attention  on  CP
differences, which  are free from background contributions  as long as
the sample environment is unchanged.  We limit ourselves to the
stated  thermodynamic  range  because  measuring the  CP  of  monomers
(coordination 0) is not possible with the required high statistics and
measuring  ice Ih (coordination  4) would  require a  different sample
environment.   We thus present our  data as  the difference  $\Delta J(p)$
between a  reference state CP (taken  as liquid water at  30 $^o$C and
300 bar) minus  the CPs at higher temperatures.  The acquisition times
are of several hours.

%
%
Our Compton scattering calculations rely on first-principles methods 
without any
structural information from experimental data (except for the density) 
\cite{density_reason}.
To  compare experiments  with theory,  we use  extensive Car-Parrinello
molecular-dynamics  simulations \cite{CPref1,CPref2,CPref3},  using 32
water molecules in periodic-boundary conditions.
The subtle technical
issues arising in the simulations  of liquid water have been discussed
extensively in the  literature \cite{GrossmanI,Schwegler2,Sit,Joost} 
-  in particular it has been  suggested \cite{rigid} that a closer agreement
with experiments  is obtained by constraining the  internal degrees of
freedom (i.e.  keeping each water  molecule rigid),  given that  at the
conditions  considered those  vibrational  degrees of  freedom are frozen out
and consequently in their  zero-point motion state. We thus  performed first-principles
molecular-dynamics   simulations   at   ambient,   subcritical   and
supercritical regimes  at 77, 200, 300,  350, 397 and  430 $^{o}$C 
\cite{thermo}
with rigid water molecules.
To determine the average water molecular structure to be used in the 
simulations mentioned above and to
understand the effects of the change in intramolecular geometries on CP
differences at various conditions, we also performed fully
unconstrained Car-Parrinello molecular dynamics simulations
\cite{flexible_ref}. These unconstrained simulations will also be used
later to further confirm the linear relationship of $n_{e}$ with $n_{HB}$.
We further choose for the
rigid-water molecular structure
the average intramolecular  O-H
distance ($r_{OH}$) and H-O-H angle ($\angle_{HOH}$) to be 0.995 \AA\
and 105.0$^{o}$ respectively, which are obtained from the 300 $^{o}$C 
simulation. 
We note as expected in our unconstrained simulations, 
a small change in average structure 
over the range of temperatures considered, with $r_{OH}$ and $\angle_{HOH}$ 
varying between
104.6 - 106.1$^{o}$ and 0.990 - 0.998 \AA\ respectively,
when the temperature decreases from
430 to 127 $^{o}$C. 
These changes have, in principle, an
effect on the CPs, but it will be shown that this effect is reduced to
that of a scale factor: the linear relation alluded to above remains robustly verified.
All our simulations were
done in the NVT ensemble,  
with the densities  fixed at  the experimental
values at 300  bar \cite{77}, using ultrasoft  pseudopotentials  with  a
plane-wave kinetic energy cutoff of  25 Ry for the wavefunctions and
200  Ry  for  the   charge  density.   The  wavefunction
fictitious mass is chosen to be 1100 a.u. and  a time step of
10  a.u. is used  to integrate  the electronic  and ionic  equations of
motions. It is important to stress that these choices  of fictitious
mass and time  step are the same as those  used in Ref.~\cite{rigid},
and are accurate for this system and wavefunction cutoff
- i.e. they provide a structure 
that is not affected by integration errors.
We note that fully-flexible water molecules described with
norm-conserving pseudopotentials would require smaller fictitious
masses ($\sim$340 a.u.) to provide accurate results - larger values
will integrate less accurately
the equations of motion, and lead to less structured radial distribution
functions as compared to tightly-controlled simulations (such choices 
improve
rather artificially the agreement with experiments, and lower the 
theoretical
freezing point of water) \cite{GrossmanI}.

%
%
The theoretical  reference state  is chosen to  be that at  77 $^{o}$C;
a  lower  temperature  would  see the  onset  of  a  freezing
transition \cite{rigid}. 
The reason for the shift of the liquid-solid
phase  boundary  have been  discussed  extensively  in the  literature
\cite{GrossmanI,Schwegler2,rigid,Sit};
increasing  the temperature  for the only case (30 $^{o}$C) close
to the liquid-solid phase transition avoids the pitfall of remaining trapped in
a glassy configuration. As
a validation  of the  theoretical temperature scale,  we found good
agreement   between  the   predicted   and  the   measured
self-diffusion coefficients. 
The predicted (and, in parenthesis, measured \cite{sd1,sd2}) self-diffusion 
coefficients 
obtained for water at 77 (30 in the experiment), 200, 300, 350, 397 
and 430 $^{o}$C are 1.5 (2.6), 9.5 (18.9), 27.0 (37.4), 35.3 (45.3), 75.9 
(80.3)
and 163.3 (174.4) $\times$10$^{-5}$cm$^{2}$/s,
respectively.
We therefore adopt liquid (rigid) water at 77 $^{o}$C, as a theoretical
counterpart for the experimental temperature of 30 $^{o}$C since the lowest
temperature for which dynamics of water remains in the liquid phase is about 
50 $^{o}$C.

%
%
To compute $J(p,{\bf e})$, we used the approach by Romero {\em et al.}
\cite{Romero},   based   on   maximally-localized  Wannier   functions
\cite{Marzari}.   Each  water simulation  lasted  about  17 ps  after
thermalization, and  at each temperature the CPs  are calculated from
the averages of 10  uncorrelated configurations, equally-spaced by 1.7
ps and then spherically averaged.   The CPs are also convoluted with a
Gaussian that matches the 0.51 a.u.  experimental resolution.

\begin{figure}
\centerline{
\rotatebox{-90}{\resizebox{2.5in}{!}{\includegraphics{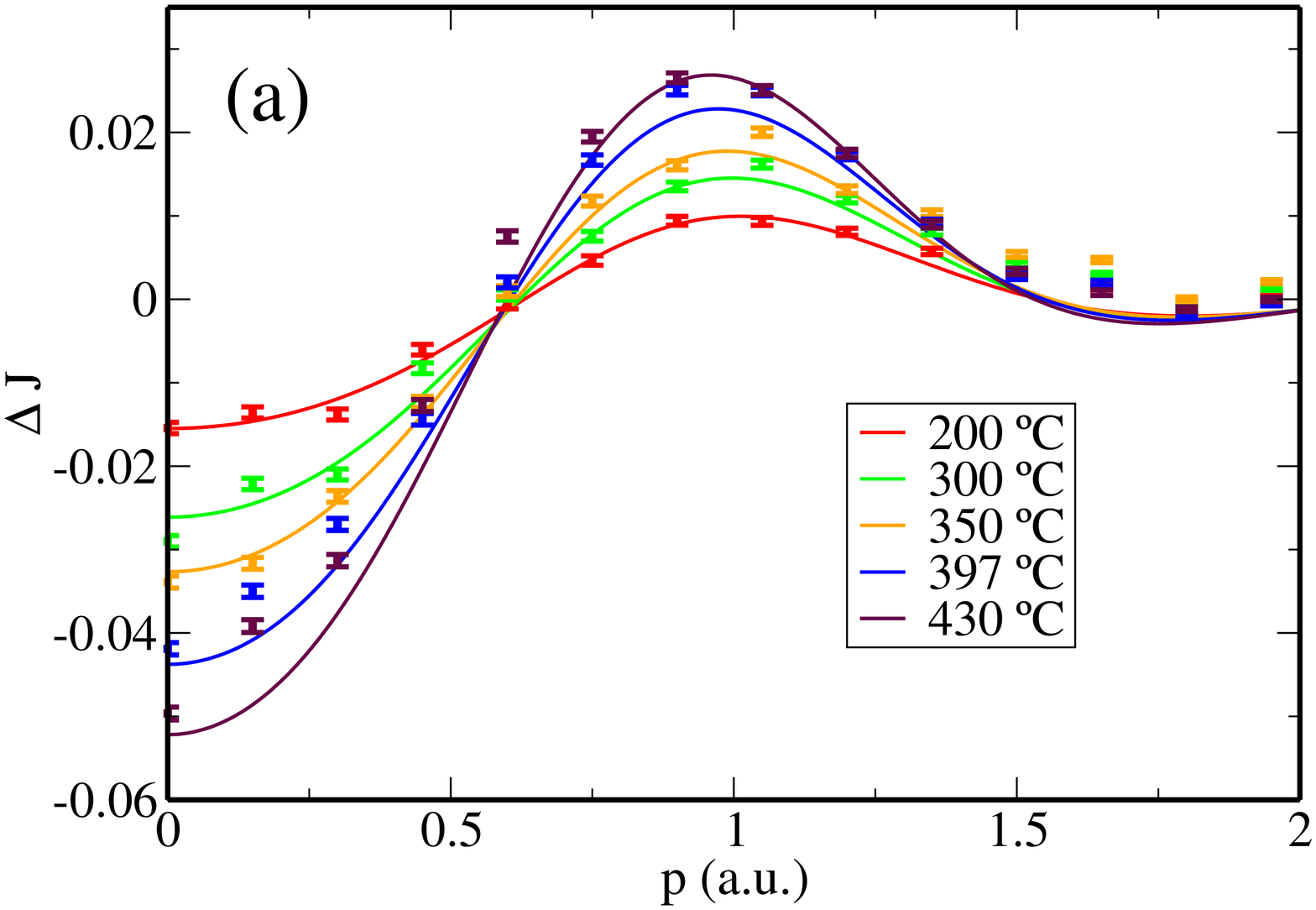}}}     }
\centerline{
\rotatebox{-90}{\resizebox{2.5in}{!}{\includegraphics{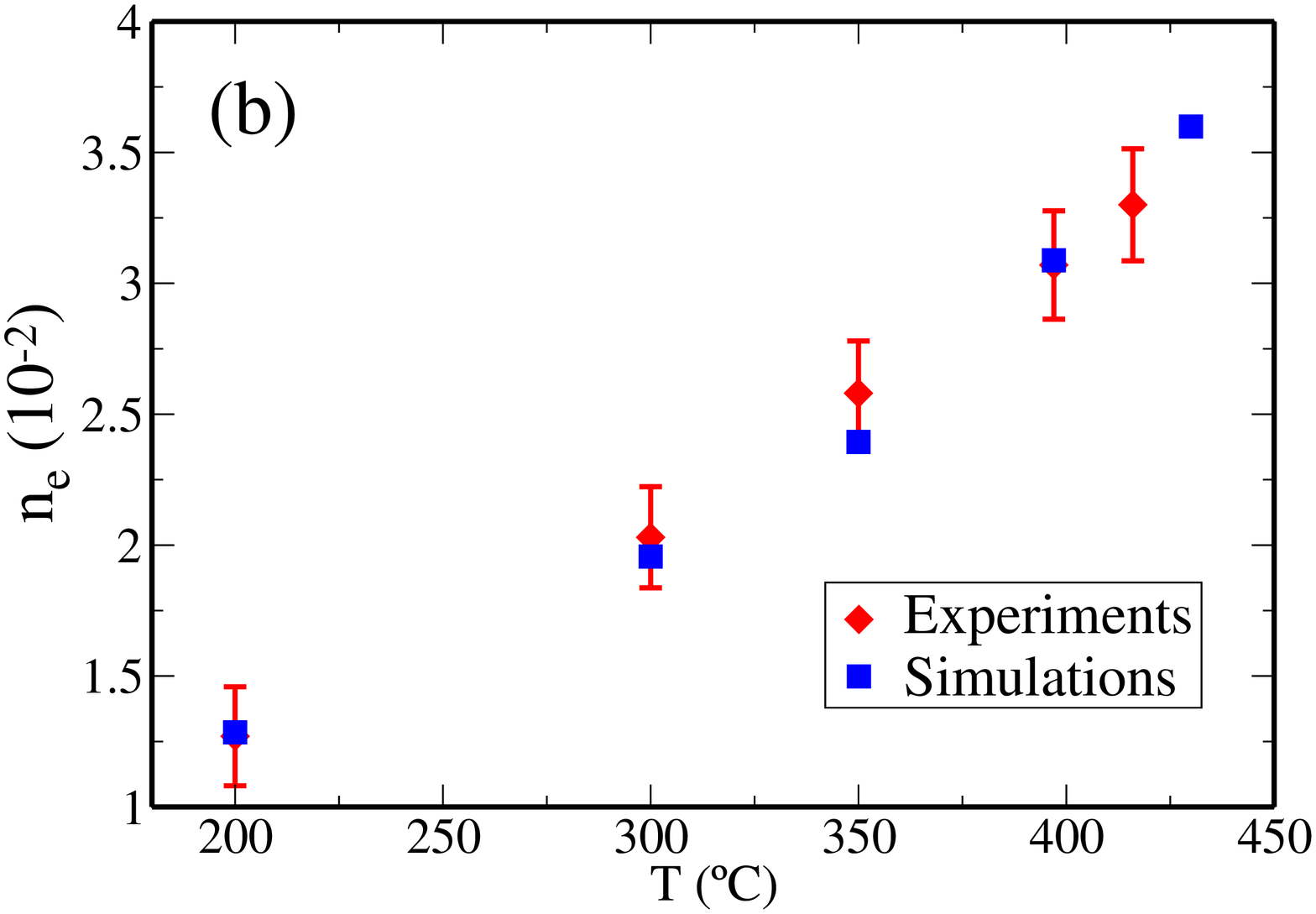}}} }
\caption{(color  online).  (a)  Experimental  CP
difference $\Delta  J(p)$ (points, in electrons  per atomic unit  of 
momentum)
with  respect to  the experimental  reference state  as a  function of
temperature, and theoretical  CP difference  $\Delta
J(p)$ (solid lines, in electrons per atomic unit of momentum) with respect 
to
the  theoretical reference state  as a  function of  temperature, and 
scaled by
0.73.  The
temperature of reference  state is 30 $^{o}$C for  the experiments, 77
$^{o}$C for the theory (at a density 0.97 g/cm$^{3}$).
For the purple experimental points the temperature is 416 $^{o}$C
while for the purple theory curve T = 430 $^{o}$C. (b) The number
of  electrons $n_{e}$  involved in  hydrogen bonding  (i.e. integrated
absolute  profile  difference)  as   a  function  of  temperature
(simulations  have
been scaled by 0.73).  }
\label{thexp}
\end{figure}

%
%
We  first  show the  experimental  and  theoretical  
$\Delta J(p)$  in Fig.~\ref{thexp}(a).   The overall  agreement between
experiment  and  theory  is  excellent, provided the theoretical curves
are rescaled to compensate for a systematic overestimation of the experimental
values. The reasons for this overestimation can be manifold; they have
been already reported in the literature \cite{Isaacs,ragot,Romero},
but will be discussed in detail later in this paper.
What should be stressed however  is  that a  {\it unique}  rescaling
factor  (here, 0.73)  brings  all   theoretical  curves  in
quantitative  agreement with  experiment.

%
%
We now propose an electronic  measure of hydrogen-bonding defined 
by the number of electrons
\begin{equation}
n_{e}= \frac{1}{2} \int_{-\infty}^{\infty}
\left|  \Delta J(p)  \right| dp, 
\end{equation}
where, as mentioned before, $\Delta J(p)$ is 
the difference between the CPs at the conditions studied and the
reference profile for normal liquid water.
The area $n_e$ (in units of electrons) between a difference 
curve and the horizontal axis in Fig.~\ref{thexp} measures the amount 
of electrons whose 
wavefunctions  change in  going from  the reference  state to  the one
being  measured.
The quantity $n_{e}$  tracks the  progressive  breaking of  hydrogen  
bonds as  the temperature is raised from ambient conditions. Both the
experimental measurements for $n_{e}$  and  the  theoretical predictions
(after  rescaling)  are plotted in Fig.~\ref{thexp}(b) 
as  a function of temperature.  All the
$n_{e}$(T)  points  line  up,  reflecting the  agreement  between  the
experimental and theoretical $\Delta J(p)$ in
Fig.~\ref{thexp}(a).  Interestingly,  $n_e$(T) in  Fig.~\ref{thexp}(b)
does not  saturate, indicating that  even at the  highest temperatures
considered significant hydrogen bonding remains.  Moreover, the almost
linear behavior  of $n_{e}$(T)  in the range  from 200 to  430 $^{o}$C
mirrors a similar trend  for $n_{HB}$, as shown in Ref.~\cite{Review1}
(Fig.~13).

\begin{figure}
\centerline{
\rotatebox{-90}{\resizebox{2.5in}{!}{\includegraphics{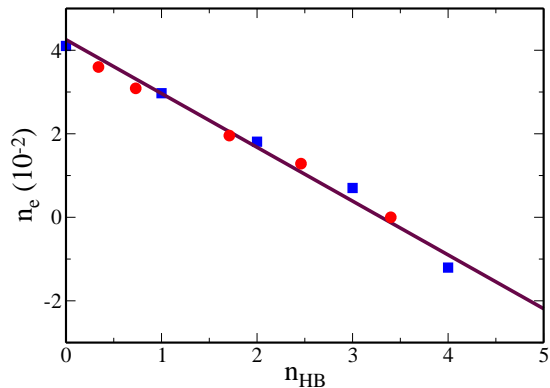}}}}
\caption{(color online).  The number of electrons  $n_{e}$ involved in
hydrogen  bonding (scaled  by 0.73)  as a  function of  hydrogen bonds
$n_{HB}$. The  zero on  the  vertical  scale  corresponds to  the
reference state at ambient conditions.  Squares: water clusters (1 to
5 molecules); circles: rigid-water bulk simulations.}
\label{nvsh}
\end{figure}

%
%
These considerations link $n_{e}$ to the variation in hydrogen bonding
as the temperature of the system is changed. In the following, we make
a direct connection  between $n_{e}$ and $n_{HB}$ by  studying a model
cluster  where  these quantities  are  unambiguous: a  five-molecule
model with one  molecule at the center and  the other four surrounding
it in a  tetrahedral arrangement.  The O-O distance  between two water
molecules in the  cluster geometry is fixed at 2.77  \AA, which is the
first peak height position  of the theoretical O-O radial distribution
function  of liquid  water at  ambient conditions within our rigid water
simulations.  It has  been shown
\cite{Hakala3} that  the CP differences are strongly  dependent on the
O-O distance. An increase in the  O-O distance will lead to a decrease
in $n_{e}$ and this can affect the scaling factor, as
we shall discuss later in the  paper.  In the present cluster model, a
single hydrogen bond  can be formed or broken  simply by removing, one
at  a time, the  four outer  molecules. This  model was  introduced in
Ref.~\cite{Hakala1},  but   here,  using  maximally-localized  Wannier
functions \cite{Marzari}, we can resolve  the CP unique to the central
molecule, as the hydrogen bonds  with its neighbors are broken, one at
a time.  The number $n_{HB}$ of  hydrogen bonds formed  by the central
molecule is  unambiguously defined, and the corresponding  CP has been
calculated by  averaging over all  the possible clusters with  a given
$n_{HB}$. $n_{e}$ can  thus be plotted as a  function of $n_{HB}$. The
results are shown in Fig.~\ref{nvsh} (squares), where $n_{e}$ is again
rescaled by  a factor of 0.73,  to facilitate a  later connection with
experiments.  A linear dependence for  the cluster results is seen and
the slope provides  an indication of the number  of electrons involved
in one hydrogen bond \cite{neg_ne}.

Finally, our first-principles  molecular dynamics  results are also  
plotted on
Fig.~\ref{nvsh},  using  for  $n_{HB}$  the structural  definition 
\cite{Geom} 
based on bond-length and bond-angle criteria.  The
$(n_e, n_{HB})$ points collapse  on the same straight line, confirming
a universal relationship between $n_e$ and $n_{HB}$, and demonstrating
that the criteria \cite{Geom} are meaningful and consistent
with  the  natural  integer  $n_{HB}$  count  of  the  cluster  model.
Remarkably,  the  linear  relationship  between $n_{e}$  and  $n_{HB}$
persists under a  wide variety of conditions, ranging  from the normal
to  the supercritical  regime  and covering  all  degrees of  hydrogen
bonding. Moreover, the slope in Fig.~\ref{nvsh}, as mentioned earlier,
provides  the number of  electrons involved  in a  hydrogen bond: 
(13 $\pm$ 0.6)$\times$10$^{-3}$  electrons are displaced for every bond 
that is broken (or  formed). 
The  bonding and  coordination picture
that emerges  from these results is  also in close  agreement with the
established understanding of coordination in liquid water. Our results
are summarized  in Table~\ref{Hbond}, where  we can see that  in going
from  the  normal  state   into  the  supercritical  regime,  $n_{HB}$
decreases   from    3.40   (compared   to   the    value   3.58   from
Ref.~\cite{soper97}) to  0.34. Moreover, except  at ambient conditions,
the theoretical $n_{HB}$'s  at different temperatures agree reasonably
well with the  NMR studies \cite{Review1}, which also  suggests that the
simulations  and experiments  produce similar  structures at  the same
temperatures above the theoretical melting point.

\begin{figure}
\rotatebox{-90}{\resizebox{2.5in}{!}{\includegraphics{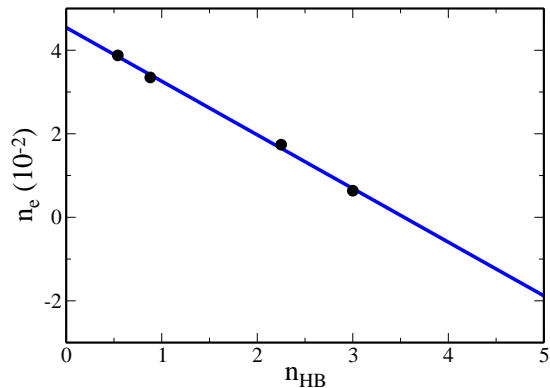}}}
\caption{ (color online). The (unscaled) 
number of electrons  $n_{e}$ involved in
hydrogen  bonding  as a  function of  hydrogen bonds
$n_{HB}$ for flexible water simulations at 200, 300, 403 and 430 $^{o}$C.
Note that the reference state in these calculations is 127 $^{o}$C due
to freezing at lower temperatures \cite{GrossmanI,Schwegler2,Sit}.}
\label{flexible}
\end{figure}

%
%
We  now tackle in detail  the origin of the  rescaling factor. In
principle, it  is possible to  exactly reproduce the amplitudes  of the
experimental  CP  differences  by  adjusting  the  intramolecular  O-H
distance     and     the    H-bond     geometry; this was detailed in
Refs.~\cite{Hakala3}~and~\cite{Nygard}.  Our approach uses instead geometries
that are obtained directly from first-principles simulations and identifies
a-posteriori
the rescaling factor that aligns experiments and simulations.
Note also that the proton zero-point
motion  is not  accounted in any these simulations and  may  have a
similar impact as discussed by Isaacs {\em et al.} \cite{Isaacs}
for the CP anisotropy of
ice (the phenomenological rescaling used in those calculations was 0.6).
For this same system, Romero {\em et al.} \cite{Romero} have
shown that finite temperature effects can be ruled out as the 
reason for rescaling and Ragot {\em et al.} \cite{ragot} demonstrated
that correlations effects are also not crucial.
It is beyond the scope of this work to determine 
whether the proton zero-point
motion or the difference in geometries due to the approximate
exchange-correlation functionals dominate the discrepancies
between experiments and theory.
Instead, we show that
the linear relation 
between $n_e$ and $n_{HB}$ holds even when intramolecular
distances are allowed to change with temperature in fully unconstrained 
simulations.
Even though these simulations are closer to the experiment 
as far as scaling 
is concerned we prefer to use the rescaled rigid 
water calculations for the fundamental reasons given above.
Fig.~\ref{nvsh} and Fig.~\ref{flexible} clearly show the remarkable
robustness of our linear relation.
Finally our fully unconstrained 
simulations clearly show that
Compton profiles may be affected by changes in internal molecular geometry, 
but these changes are caused by hydrogen bonding and cannot be treated 
in a separate way.

\begin{figure}
\hspace{1.2in}
\rotatebox{0}{\resizebox{5in}{!}{\includegraphics{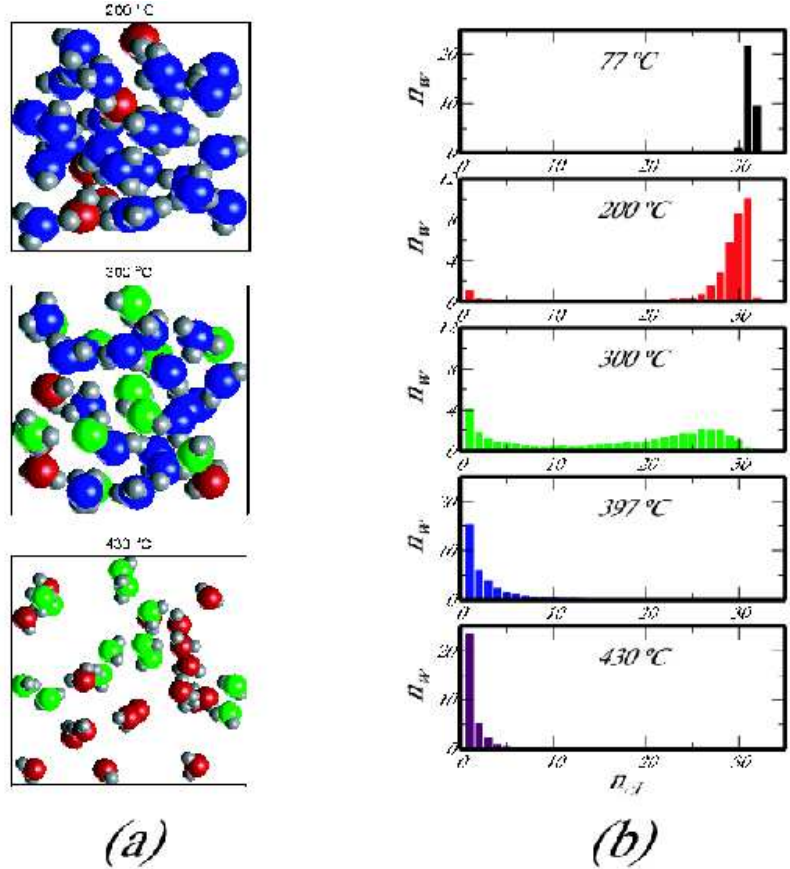}}}
\caption{(color online).   (a) Snapshots  from the simulations  at 200
$^{o}$C,  300  $^{o}$C and  430  $^{o}$C.   Each  panel shows  the  32
inequivalent molecules in the  periodic cubic supercell (note that the
volumes  at  these conditions  are  very  different,  and the  figures
presented here  have been rescaled accordingly).  The  color codes the
connectivity of each water  molecule (see text). Blue: water molecules
in  clusters with  6  or  more molecules;  green:  water molecules  in
clusters with  2 to 5 molecules;  red: monomers.  (b)  Number of water
molecules  $n_w$  that  belong  to  a cluster  of  size  $n_{cl}$,  at
different temperatures.  }
\label{mer}
\end{figure}

%
%
We now  examine the  connectivity of the  hydrogen-bond network  as we
move from the  normal to the supercritical regime.  For each molecule,
we can determine the number  of neighbors on a connected hydrogen-bond
network.   Fig.~\ref{mer}(a)  shows  three  snapshots taken  from  our
rigid water simulations.  We see  that at 200 $^{o}$C most  molecules (in 
blue) are
in clusters containing 6 or  more molecules, and very few monomers (in
red)  exist. Most  remarkably, no  intermediate cluster  size  exist -
i.e.  a molecule  is either  fully connected  to all  other,  or fully
disconnected, probably reminiscent of the very high surface tension of
water.  At 300 $^{o}$C, there is a very  wide distribution of cluster
sizes: monomers  and small  clusters containing 2  to 5  molecules (in
green).  Finally, at 430 $^{o}$C monomers dominate, but a few clusters
of less than  6 molecules persist. To quantify  these observations, we
show in Fig.~\ref{mer}(b) the distribution of cluster sizes at a given
simulation - $n_{w}$ molecules belong  to a cluster (of size $n_{cl}$)
if  they  are  hydrogen  bonded  together.   The  uppermost  panel  of
Fig.~\ref{mer}(b)  shows that  at normal  conditions almost  all water
molecules are  connected to each other.  The percolating hydrogen-bond
network  breaks somewhere  between  300 $^{o}$C  and  397 $^{o}$C,  as
illustrated   in  Fig.~\ref{mer}(b)   by  the   diffuse   and  bimodal
distribution at  300 $^{o}$C and the  peaked, monotonically decreasing
distribution at 397 $^{o}$C.   According to Table~\ref{Hbond}, this is
consistent     with    the     well-known     percolation    threshold
$n_{HB}=1.53-1.55$  \cite{stanley}.   Finally,  the  lowest  panel  in
Fig~\ref{mer}(b) shows a distribution at 430 $^{o}$C with a dominating
amount  of  monomers but  also  a  substantial  amount of  dimers  and
trimers,  indicating  that  the  amount  of hydrogen  bonds  is  still
non-negligible  in  supercritical  conditions. The  experiment  yields
$n_{HB}$=0.74$\pm$0.16 at  416 $^{o}$C, while  the simulation predicts
$n_{HB}$=0.34 at 430 $^{o}$C. These values are consistent with Fig. 13
in the review article by Kalinichev \cite{Review1}.

\begin{table}
\caption{  Experimental and  theoretical (scaled  by 0.73)  $n_e$, and
theoretical $n_{HB}$ at different state points. $n_{HB}$ is calculated
from  the   molecular  dynamics  trajectories   using  the  structural
criterion  of   Ref.~\cite{Science}.   The  first   column  shows  the
experimental  temperature  and  pressure  at  each  state  point.}
\begin{ruledtabular}
\begin{tabular}{c|c|c|c}
  Temperature &  $n_{e}$ (10$^{-2}$) & $n_{e}$  (10$^{-2}$) & $n_{HB}$
\\  and pressure &  (experiment) &  (theory) &  (theory) \\  \hline 
200 $^{o}$C, 300 bar & 1.27 $\pm$ 0.19
& 1.28 & 2.46 \\ 300 $^{o}$C, 300  bar & 2.03 $\pm$ 0.19 & 1.96 & 1.71
\\  350 $^{o}$C,  300 bar  & 2.58  $\pm$  0.20 &  2.39 &  1.38 \\  397
$^{o}$C, 300 bar &  3.07 $\pm$ 0.21 & 3.09 & 0.73  \\ 416 $^{o}$C, 300
bar & 3.30 $\pm$ 0.21 &  &  \\
430 $^{o}$C, 300
bar &  & 3.60 & 0.34 \\

\end{tabular}
\end{ruledtabular}
\label{Hbond}
\end{table}

%
%
In conclusion, we have  shown that Compton scattering provides an
absolute measure of the number $n_{e}$ of electrons per water molecule
involved in  hydrogen bonds. We  have measured the dependence of  $n_{e}$ 
as
a function of
temperature and compared  with  state-of-the-art  
first-principles
molecular  dynamics  simulations. 
This comparison leads  to  
a  linear relation  between
$n_{e}$ and  the average number  $n_{HB}$ of hydrogen bonds  per water
molecule. The linear relation is very robust and holds  over 
all possible coordination numbers.
This approach  should prove valuable  in the future to  address issues
that arise in the characterization of hydrogen-bonded networks.

We acknowledge  useful discussions with  G. Loupias and  V. Honkimaki,
and  thank B. Wood  for his  help and  suggestions in  calculating the
connectivity of  a bonding network.  We also  acknowledge support from
the  Croucher  Foundation (P.H.-L.S.),  MURI  grant DAAD  19-03-1-0169
(N.M.), the MRSEC Program of the National Science Foundation under the
award   number    DMR   02-13282   (P.H.-L.S.)    and   
U.S.D.O.E. contracts DE-AC03-76SF00098,  
DE-FG02-07ER46352 (B.B.).  The  calculations in  this work  have been
performed   using   the   Quantum-ESPRESSO  package   \cite{ESPRESSO}.
Computational  facilities   have  been  provided   through  NSF  grant
DMR-0414849 and  PNNL grant EMSL-UP-9597 at MIT,  and the Northeastern
University's Advanced Scientific Computation Center (ASCC).


\begin{thebibliography}{10}
\bibitem{density_reason}
While also the density could be determined fully from first-principles
constant pressure simulations, the timescale needed for statistical
accuracy are still beyond ordinary computer resources, especially in the
supercritical regimes.
\bibitem{thermo}
In all the molecular dynamics simulations, a Nose-Hoover thermostat is
applied to the ions, but no electronic thermostat is used.
\bibitem{flexible_ref}
The details of simulations at ambient conditions can be found in 
Ref.~\cite{Sit}.
We also performed simulations at 200, 300, 403 and 430 $^{o}$C, all at
300 bar. Pseudopotentials, fictitious mass and timestep are the same
as those used in Ref.~\cite{Sit}. Unlike the rigid water simulations,
all the flexible water simulations were done with heavy water molecules.
\bibitem{77} At 77 $^{o}$C, the density  is taken to be that at 1 bar,
and differs by less than 4\% from the one at 300 bar.
\bibitem{neg_ne} The  negative value of $n_{e}$ for  $n_{HB}$=4 is due
to the fact that the definition of $n_{e}$ should be generalized, when
the reference state is less  coordinated than the considered state, to
$n_{e}=       \frac{-\Delta      J(0)}{2\left|\Delta      J(0)\right|}
\int_{-\infty}^{\infty} \left| \Delta J(p) \right| dp$.
\bibitem{Geom} We adopt the geometrical  criterion for H-bond
suggested in Ref.~\cite{Science}. It
is  considered  to  be  a  hydrogen  bond  when  $r_{OO}$  $\leq$  3.3
\AA\ -0.00044$\theta^{2}$,   where  $\theta$   is   the  H-bond   O-H-O
angle. $\theta$=0 when the H-bond is linear.
\end{thebibliography}
\end{document}